\documentclass[a4paper]{article}

\usepackage{INTERSPEECH2022}

\usepackage{multirow}
\usepackage[table,xcdraw]{xcolor}
\usepackage{hyperref}
\hypersetup{
    colorlinks=true,
    linkcolor=blue,
    urlcolor=blue,
    }
\newcommand{\vect}[1]{\boldsymbol{\mathrm{#1}}}

\newcommand{\norm}[1]{\left\lVert#1\right\rVert}

\title{Robust Speaker Recognition with Transformers Using wav2vec 2.0}
\name{Sergey Novoselov$^{1,2}$,  Galina Lavrentyeva$^{1,2}$, Anastasia Avdeeva$^{1,2}$, Vladimir Volokhov$^{1,2}$, \\Aleksei Gusev$^{1,2}$}
\address{
$^1$ STC Ltd., St. Petersburg, Russia \\
$^2$ ITMO University, St. Petersburg, Russia}
\email{\{novoselov,  lavrentyeva, avdeeva-a, volokhov gusev-a\}@speechpro.com}

\begin{document}

\maketitle
\begin{abstract}
Recent advances in unsupervised speech representation learning discover new approaches and provide new state-of-the-art for diverse types of speech processing tasks. This paper presents an investigation of using wav2vec 2.0 deep speech representations for the speaker recognition task. The proposed fine-tuning procedure of wav2vec 2.0 with simple TDNN and statistic pooling back-end using additive angular margin loss allows to obtain deep speaker embedding extractor that is well-generalized across different domains. It is concluded that Contrastive Predictive Coding pretraining scheme efficiently utilizes the power of unlabeled data, and thus opens the door to powerful transformer-based speaker recognition systems. The experimental results obtained in this study demonstrate that fine-tuning can be done on relatively small sets and a clean version of data. Using data augmentation during fine-tuning provides additional performance gains in speaker verification. In this study speaker recognition systems were analyzed on a wide range of well-known verification protocols: VoxCeleb1 cleaned test set, NIST SRE 18 development set, NIST SRE 2016 and NIST SRE 2019 evaluation set,  VOiCES evaluation set, NIST 2021 SRE, and CTS challenges sets.
\end{abstract}

\section{Introduction}

\label{sec:intro}

Today's state-of-the-art \cite{Zeinali2019, garcia2020magneto, gusev2020deep, lee2019nec, evalplan2021nist} speaker recognition (SR) systems are based on very deep convolutional neural networks (\textit{ResNets}, \textit{ECAPA-TDNNs}) taking log Mel Filter Bank features as input and that are trained on large datasets using additive angular margin loss functions and different optimization strategies. The simple cosine or PLDA scorings are usually used as extractor back-ends.  

Recent advances in unsupervised speech representation learning \cite{oord2018representation, schneider2019wav2vec, baevski2020wav2vec, chen2021wavlm, hsu2021hubert} discover new approaches and provide new state-of-the-art models for diverse types of speech processing tasks including speaker recognition. The key goal of such models is speech prediction modeling \cite{oord2018representation} or speech denoising modeling \cite{chen2021wavlm} which can be done in an unsupervised manner. For example, the reports \cite{chen2021wavlm} about strong results of \textit{WavLM} model fine-tuning for speech recognition, speaker recognition, speech separation, and speaker diarization tasks. It should be noted that an important aspect of such models is the utilization of a powerful transformer structure \cite{baevski2020wav2vec} as the backbone model which takes raw speech signals as input and incorporates a multi-head attention mechanism on the deep layers.

% In contrast to past NIST SREs \cite{sadjadi20172016, sadjadi20202019, matvejka202013} the key challenge provided by new NIST SRE 21 datasets \cite{evalplan2021nist} are multi-channel and multi language  speaker recognition based on audio-from video and telephone speech segments. So the top performing systems should be well calibrated and robust for the cross-channel and same-channel microphone and telephone conditions. To this end in our work we considered both 8 kHz and 16 kHz acoustic features to train different system.
The authors of \cite{fan2020w2v_speaker, vaessen2021fine} share good results of their attempts to fine-tune \textit{wav2vec 2.0} model for VoxCeleb \cite{nagrani2017voxceleb, chung2018voxceleb2} speaker recognition sets.

Inspired by the success of wav2vec 2.0 in speech recognition tasks \cite{schneider2019wav2vec, baevski2020wav2vec} in our recent \cite{avdeeva2021stc} and current works we performed a new study of wav2vec 2.0 model fine-tuning for speaker recognition tasks. We used large multi-lingual wav2vec 2.0 models provided by Facebook \cite{ott2019fairseq} in the
fairseq repository\footnote{\href{https://github.com/pytorch/fairseq/tree/main/examples/wav2vec}{https://github.com/pytorch/fairseq/tree/main/examples/wav2vec}}
% \href{https://github.com/pytorch/fairseq/tree/main/examples/wav2vec}{\textit{fairseq cite}}
as a starting point of our fine-tuning. In the wav2vec 2.0 based speaker recognition encoder network, we proposed to use TDNN and statistic pooling layers based back-end. Additionally, we explored the questions of optimal transformers layer selection and usefulness of audio augmentation during model fine-tuning for speaker recognition.

% It should be noted that we started  wav2vec 2.0 based SR model investigation during our participation \cite{avdeeva2021stc} in the NIST SRE 2020-2021 CTS challenges \cite{sadjadi2020nist}. 

As baseline speaker recognition systems in this work we utilized ResNets and ECAPA-TDNN models and their fusion.
The experiments were conducted on a wide range of well-known verification benchmarks described in Section \ref{ssec:test_set}.

\section{Speaker recognition systems}
\label{sec:sr_sys}
The conventional deep neural network based solution for extracting utterance-level speaker embeddings consists of three blocks: an encoder network for extracting frame-level representations from the acoustic features, pooling layer that converts variable-length frame-level features into one fixed-dimensional vector and a feed forward classification network that processes the pooling vector to produce speaker class posterior. 
%During the evaluation step 

The role of the encoder network can be taken by a neural network of any type. We aimed to explore state-of-the-art architectures in speaker recognition and related fields for this purpose: we considered ResNet and TDNN based architectures (Section \ref{ssec:baseline_sys}) as our baseline systems. They have already shown impressive performance in the speaker verification domain. Alternative  transformer-based approaches like wav2vec 2.0 model fine-tuning are described in Section \ref{ssec:w2v_sys}.

Several papers confirm \cite{garcia2020magneto,gusev2020deep,garcia2019x} the effectiveness of the training scheme where neural networks that are first trained on short utterances are then fine-tuned using longer utterances. We followed this approach during this study and first trained extractors on 4-6 sec speech chunks and then fine-tuned on 12-18 sec segments.

According to our experience considered deep speaker embedding extractors contain huge amounts of trainable parameters and are capable enough to solve the speaker recognition task without complex back-end or preprocessing steps. It can be trained to perform all necessary calculations by itself, given sufficient amounts of diverse training data.
Following this intuition during our experiments we were mainly focused on training powerful deep speaker embedding extractors and didn't pay much attention to its front-end and back-end.
% In order to deal with cross-channel conditions we considered both 8 kHz and 16 kHz acoustic features to train different system.

For training and tuning processes of all our extractors the additive angular margin softmax (AAM-Softmax) based loss was used with parameters $m$ and $s$ set to 0.35 and 32 respectively.

We used the one cycle learning rate policy \cite{onecycle} in all our experiments.

\subsection{Baseline systems}
\label{ssec:baseline_sys}
\subsubsection{Front-End processing}
\label{sssec:base_front-end}
%First we would like to describe front-end processing and low level features used in our speaker recognition systems. 

In this research Log Mel-filter bank features (MFB) were used as low-level features for the baseline systems:
\begin{itemize}
\item \textbf{8kHz features} extracted from raw audio signal (8000 Hz) with 25ms frame-length and 10 ms overlap. The frequency coverage was from 20 Hz to 3700 Hz with the number of mel bins 64.
\end{itemize}

Mean Normalization (MN) over a 3-second sliding window was applied after the features were extracted.  

U-net-based VAD \cite{gusev2020deep} was used to filter out non-speech frames. 

\subsubsection{ResNet101 encoder network}
Proposed in 2015 for computer vision task, ResNet \cite{resnet} is now one of the most popular architectures. By introducing the shortcut connections to the CNN, the ResNet model is able to build very deep neural networks and achieve remarkable performance in speaker recognition under challenging conditions \cite{but_resnet, jhu_resnet}.
This network uses 2-dimensional features as input and processes them using 2-dimensional convolution in both the time and frequency domains.

We used ResNet101 as a baseline model. The model configuration was the following:
\begin{itemize}
    \setlength\itemsep{-0.0em}
    \item Simple Conv2D layer in place of stem block;
    \item BottleneckBlock blocks;
    \item Number of filters [32, 64, 128, 256];
    \item Stride = 1 in BottleneckBlock;
    \item Maxout activation function for the embedding layer;
    \item Embedding size - 512.
\end{itemize}

\subsubsection{ECAPA-TDNN encoder network}

Enhanced TDNN architecture with Emphasized channel attention, propagation and aggregation, proposed in \cite{desplanques2020ecapa}, is a modification of the standard time delay neural network (TDNN) architecture, containing squeeze-excitation (SE) blocks and Res2Net modules in the frame level with hierarchical filters for the extraction of features of different scales. To process signals of arbitrary duration, the architecture uses attentive statistic pooling instead of the conventional statistic pooling.

We used ECAPA-TDNN as our second baseline extractor. In our implementation of this model, adaptive statistic pooling and 4 SE-Res2Net Blocks with dilation values 2,3,4,5 were used. The model configuration was as follows:
\begin{itemize}
    \setlength\itemsep{-0.0em}
    \item 1024 filters in convolution frame layers;
    \item Stem block: the stack of 4 Conv2D-BatchNorm2D-ReLU sequences with kernel  size 3 and 32 filters for all  convolution  layers  except the last one that used 1024 filters;
    \item Maxout activation function for the embedding layer;
    \item Embedding size - 512.
\end{itemize}

\subsubsection{Scoring}
We used cosine similarity to distinguish speaker embeddings:
\begin{equation}
\label{eq:cos}
\mathcal{S(\vect{e_1},\vect{e_2})} = \dfrac{\vect{e_1}^T\vect{e_2}}{{\norm{\vect{e_1}}}{\norm{\vect{e_2}}}},
\end{equation}
where $(\vect{e_1}, \vect{e_2})$ are classification layer speaker embedding (cl-embeddings) vectors or class posteriors logit embeddings\cite{avdeeva2021stc}.

Additionally, adaptive score normalization technique (adaptive s-norm) from \cite{CVDFKCL2017} was used.

In \cite{avdeeva2021stc} we found out that using channel normalization is effective for cross-channel evaluations. According to this, in our experiments on cross-channel protocols from  NIST SRE2021 we used the same technique. The normalized score for a pair $(\vect{e_1},\vect{e_2})$ can be estimated as follows:
\begin{equation}
\label{eq:score_norm2}
\mathcal{\hat{S}(\vect{e_1},\vect{e_2})} = \frac{\mathcal{S(\vect{e_1},\vect{e_2})}-\mu_{ch}}{\sigma_{ch}},
\end{equation}
where mean $\mu_{ch}$ and standard deviation $\sigma_{ch}$ are calculated for each pair of source type matching (tel-tel, mic-mic, tel-mic and mic-tel), obtained from source files headers and applied according to the source type of $(\vect{e_1},\vect{e_2})$ \cite{avdeeva2021stc}.

\subsection{Wav2vec 2.0 based system}
\label{ssec:w2v_sys}
\subsubsection{Front-End processing}
\label{sssec:w2v_front-end}
\textbf{Raw audio signal} (16 kHz audio) was used for our wav2vec 2.0 based extractors. 

Similarly to \ref{ssec:baseline_sys} systems U-net-based VAD \cite{gusev2020deep} was used to filter out non-speech frames.

\subsubsection{Wav2vec-TDNN encoder network}
\textbf{Wav2vec 2.0} model is a powerful transformer-based model developed for ASR tasks. It takes raw speech signals as input and incorporates a multi-head attention mechanism on the deep layers. The key aspect of training such a model is Contrastive Predicting Coding \cite{oord2018representation} self-supervised pretraining scheme. It was shown in \cite{baevski2020wav2vec} that wav2vec 2.0 model pretrained on large amounts of diverse and unlabelled data can be successfully fine-tuned to specific low resource ASR tasks. 

\begin{figure}[t!]
\centering
\includegraphics[scale=0.8]{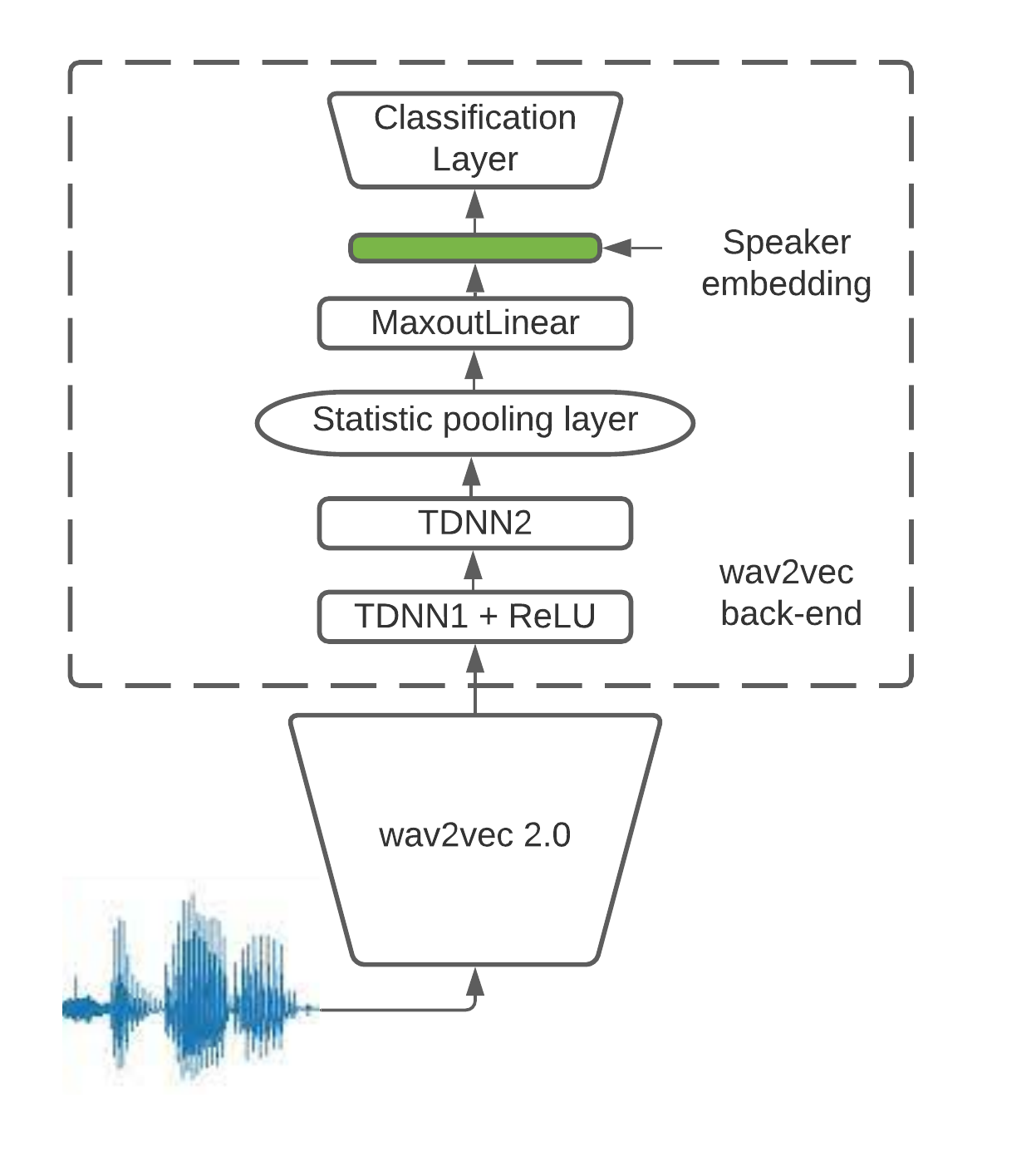}
\caption{Wav2vec 2.0 based speaker embeddings extractor}
\label{fig:w2v_arch}
\end{figure}

The proposed main scheme of wav2vec 2.0 based speaker embeddings extractor is presented in Figure \ref{fig:w2v_arch}.
As an effective wav2vec 2.0 backend we applied two TDNN layers (the 1st with ReLU activation), statistic pooling layer to pool time series to a single vector, maxout linear layer \cite{gusev2020deep, novoselov2018deep} to obtain speaker embedding. We named such a model wav2vec-TDNN. We used an AAM-Softmax-based linear classification layer to fine-tune the extractor. In principle, one can pass the output of the wav2vec 2.0 directly to the statistics pooling layer \cite{fan2020w2v_speaker}. However, we found out that we can achieve better results if we first pass them through the sequence of TDNN layers. The role of TDNN layers is to prefilter speaker-specific information and to "prepare" wav2vec 2.0 output time series for statistical pooling. The TDNN blocks utilize context 1 of the input features and have 2048-dimensional outputs. The obtained final speaker embedding size was 512. One additional point of interest is that wav2vec part of the extractor could be frozen while tuning for the downstream speaker recognition task. We observed that in this scenario the results can also be very impressive, but fine-tuning the whole extractor provides additional performance gains for the speaker recognition system.

Considered wav2vec-TDNN models were based on wav2vec 2.0 large architectures.
Large multi-lingual wav2vec 2.0 models like \textit{XLSR\_53}\footnotemark[1] and \textit{XLS-R\_1B} \footnotemark[2] provided by Facebook \cite{ott2019fairseq, babu2021xlsr} were used as starting points for the fine-tuning. We named the corresponding speaker embedding extractors as wav2vec-TDNN(\textit{XLSR\_53}) and wav2vec-TDNN(\textit{XLS-R\_1B}) respectively.

\footnotetext[1]{\href{https://github.com/pytorch/fairseq/tree/main/examples/wav2vec}{https://github.com/pytorch/fairseq/tree/main/examples/wav2vec}}
\footnotetext[2]{\href{https://github.com/pytorch/fairseq/tree/main/examples/wav2vec/xlsr}{https://github.com/pytorch/fairseq/tree/main/examples/wav2vec/xlsr}}

\subsubsection{Scoring}
Similarly to baseline systems (Section \ref{ssec:baseline_sys}) for wav2vec-TDNN we applied cosine similarity scoring (\ref{eq:cos}) in cl-embeddings space with adaptive s-norm and channel normalization postprocessing.

\section{Experimental setup}
\subsection{Train datasets}
\label{ssec:train dataset}
A wide variety of different datasets containing telephone and microphone data from proprietary datasets and from those available online were used to train the SR systems:
\begin{itemize}
    \setlength\itemsep{-0.3em}
    \item Switchboard2 Phases 1, 2 and 3;
    \item Switchboard Cellular;
    \item Mixer 6 Speech;
    \item NIST SREs 2004 - 2010;
    \item NIST SRE 2018 (eval set);
    \item concatenated VoxCeleb (VC) 1 and 2;
    \item RusTelecom v2; %extended versions of the Russian speech subcorpus named RusTelecom v2
    \item RusIVR corpus.
\end{itemize}
RusTelecom v2 is an extended version of a private Russian corpus of telephone speech, collected by call centers in Russia. 
RusIVR is a private Russian corpus with telephone and media data, collected in various scenarios and recorded using different types of devices (telephone, headset, far-field microphone, etc).
% In order to increase the amount and diversity of the training data, we used Kaldi standard recipe in addition to augmentations described in section \ref{Fix train datasets}.
In total, this training set contains 532,541 records from 33,466 speakers.

\subsubsection{Augmentations}
\label{sssec:augments}
% In order to increase the amount and diversity of the training data 
For the baseline systems, we utilized standard Kaldi augmentation recipe (reverberation, babble, music and noise) with freely available MUSAN and simulated Room Impulse Response (RIR) datasets.

In the case of wav2vec 2.0 based system tuning online augmentation scheme was used for raw audio samples with the following settings:
\begin{itemize}
    \setlength\itemsep{-0.3em}
    \item MUSAN additive noise with $p=0.25$;
    \item RIR convolution with $p=0.25$;
    \item Frequency masking with $p=0.25$;
    \item Time masking with $p=0.25$;
    \item Clipping distortion with $p=0.25$.
\end{itemize}
Here $p$ is a probability of applying augmentation type for the sample in the training batch. All considered augmentations were applied in sequence.

\begin{table}[h]
\caption{Results of speaker verification on VC1-O (cleaned) and SRE'18 dev sets in dependence of wav2vec 2.0 encoder output layer selection for wav2vec-TDNN(\textit{XLSR\_53})$^3$}
\label{tab:w2v-tdnnclean}
\resizebox{\columnwidth}{!}{
\begin{tabular}{|c|c|cc|cc|}
\hline
\rowcolor[HTML]{FFFFC7} 
\cellcolor[HTML]{FFFFC7}                                           & \cellcolor[HTML]{FFFFC7}{\color[HTML]{172B4D} }                                     & \multicolumn{2}{c|}{\cellcolor[HTML]{FFFFC7}\textbf{VC1-O (cleaned)}} & \multicolumn{2}{c|}{\cellcolor[HTML]{FFFFC7}\textbf{SRE'18   dev}}         \\ \cline{3-6} 
\rowcolor[HTML]{FFFFC7} 
\multirow{-2}{*}{\cellcolor[HTML]{FFFFC7}\textbf{Layer}} & \multirow{-2}{*}{\cellcolor[HTML]{FFFFC7}{\color[HTML]{172B4D} \textbf{Train set}}} & \multicolumn{1}{c|}{\cellcolor[HTML]{FFFFC7}\textbf{EER}} & \textbf{DCF(0.01)} & \multicolumn{1}{c|}{\cellcolor[HTML]{FFFFC7}\textbf{EER}} & \textbf{DCF(0.01)} \\ \hline
3                                                                  & {\color[HTML]{172B4D} }                                                             & \multicolumn{1}{c|}{2.54}                                 & 0.29               & \multicolumn{1}{c|}{13.58}                                & 0.62               \\ \cline{1-1} \cline{3-6} 
6                                                                  & {\color[HTML]{172B4D} }                                                             & \multicolumn{1}{c|}{1.82}                                 & 0.22               & \multicolumn{1}{c|}{10.19}                                & 0.51               \\ \cline{1-1} \cline{3-6} 
9                                                                  & {\color[HTML]{172B4D} }                                                             & \multicolumn{1}{c|}{1.76}                                 & 0.197              & \multicolumn{1}{c|}{10.5}                                 & 0.48               \\ \cline{1-1} \cline{3-6} 
12                                                                 & {\color[HTML]{172B4D} }                                                             & \multicolumn{1}{c|}{1.71}                                 & 0.21               & \multicolumn{1}{c|}{10.58}                                & 0.52               \\ \cline{1-1} \cline{3-6} 
18                                                                 & {\color[HTML]{172B4D} }                                                             & \multicolumn{1}{c|}{\textbf{1.61}}                        & \textbf{0.17}      & \multicolumn{1}{c|}{\textbf{9.97}}                        & \textbf{0.44}      \\ \cline{1-1} \cline{3-6} 
24                                                                 & \multirow{-6}{*}{{\color[HTML]{172B4D} \begin{tabular}[c]{@{}l@{}} VC1 \end{tabular}}}                                  & \multicolumn{1}{c|}{na\footnotemark[4]}                                   & na                 & \multicolumn{1}{c|}{na}                                   & na                 \\ \hline
\end{tabular}
}
\end{table}

\begin{table}[h]
\caption{Results of speaker verification on VC1-O (cleaned) test and SRE'18 dev sets in dependence of wav2vec 2.0 encoder output layer selection for wav2vec-TDNN(\textit{XLSR\_53})$^3$}
\label{tab:w2v-tdnnaug}
\resizebox{\columnwidth}{!}{
\begin{tabular}{|c|c|cc|cc|}
\hline
\rowcolor[HTML]{FFFFC7} 
\cellcolor[HTML]{FFFFC7}                                           & \cellcolor[HTML]{FFFFC7}{\color[HTML]{172B4D} }                                     & \multicolumn{2}{c|}{\cellcolor[HTML]{FFFFC7}\textbf{VC1-O (cleaned)}} & \multicolumn{2}{c|}{\cellcolor[HTML]{FFFFC7}\textbf{SRE'18  dev}}         \\ \cline{3-6} 
\rowcolor[HTML]{FFFFC7} 
\multirow{-2}{*}{\cellcolor[HTML]{FFFFC7}\textbf{Layer}} & \multirow{-2}{*}{\cellcolor[HTML]{FFFFC7}{\color[HTML]{172B4D} \textbf{Train set}}} & \multicolumn{1}{c|}{\cellcolor[HTML]{FFFFC7}\textbf{EER}} & \textbf{DCF(0.01)} & \multicolumn{1}{c|}{\cellcolor[HTML]{FFFFC7}\textbf{EER}} & \textbf{DCF(0.01)} \\ \hline
3                                                                  & {\color[HTML]{172B4D} }                                                             & \multicolumn{1}{c|}{3.47}                                 & 0.327              & \multicolumn{1}{c|}{12.22}                                & 0.55               \\ \cline{1-1} \cline{3-6} 
6                                                                  & {\color[HTML]{172B4D} }                                                             & \multicolumn{1}{c|}{2.37}                                 & \textbf{0.227}     & \multicolumn{1}{c|}{\textbf{9.78}}                        & \textbf{0.45}      \\ \cline{1-1} \cline{3-6} 
9                                                                  & {\color[HTML]{172B4D} }                                                             & \multicolumn{1}{c|}{2.23}                                 & 0.267              & \multicolumn{1}{c|}{10.88}                                & 0.48               \\ \cline{1-1} \cline{3-6} 
12                                                                 & {\color[HTML]{172B4D} }                                                             & \multicolumn{1}{c|}{2.38}                                 & 0.321              & \multicolumn{1}{c|}{10.34}                                & 0.45               \\ \cline{1-1} \cline{3-6} 
18                                                                 & {\color[HTML]{172B4D} }                                                             & \multicolumn{1}{c|}{\textbf{2.21}}                        & 0.243              & \multicolumn{1}{c|}{11.06}                                & 0.54               \\ \cline{1-1} \cline{3-6} 
24                                                                 & \multirow{-6}{*}{{\color[HTML]{172B4D} \begin{tabular}[c]{@{}l@{}} VC1 \\ + augs \end{tabular}}}                                  & \multicolumn{1}{c|}{16.62}                                & 0.99               & \multicolumn{1}{c|}{30}                                   & 1                  \\ \hline
\end{tabular}
}
\end{table}
\footnotetext[3]{Model tuned during 20 epochs}
\footnotetext[4]{No convergence achieved}

\begin{table}[h]
\caption{Results of speaker verification on VC1-O (cleaned) test and SRE'18 development sets for wav2vec-TDNN(\textit{XLSR\_53})$^3$}
\label{tab:w2v-aug}
\resizebox{\columnwidth}{!}{
\begin{tabular}{|c|cc|cc|}
\hline
\rowcolor[HTML]{FFFFC7} 
 \cellcolor[HTML]{FFFFC7}{\color[HTML]{172B4D} }                                     & \multicolumn{2}{c|}{\cellcolor[HTML]{FFFFC7}\textbf{VC1-O (cleaned)}} & \multicolumn{2}{c|}{\cellcolor[HTML]{FFFFC7}\textbf{SRE'18  dev}}         \\ \cline{2-5} 
\rowcolor[HTML]{FFFFC7} 
 \multirow{-2}{*}{\cellcolor[HTML]{FFFFC7}{\color[HTML]{172B4D} \textbf{Train set}}} & \multicolumn{1}{c|}{\cellcolor[HTML]{FFFFC7}\textbf{EER}} & \textbf{DCF(0.01)} & \multicolumn{1}{c|}{\cellcolor[HTML]{FFFFC7}\textbf{EER}} & \textbf{DCF(0.01)} \\ \hline
{\color[HTML]{172B4D} \begin{tabular}[c]{@{}l@{}} VC1+VC2\end{tabular}}                                        & \multicolumn{1}{c|}{0.86}                                 & 0.082              & \multicolumn{1}{c|}{9.07}                                 & 0.47               \\ \hline
 {\color[HTML]{172B4D} \begin{tabular}[c]{@{}l@{}} VC1+VC2 \\ + augs \end{tabular} }                         & \multicolumn{1}{c|}{\textbf{0.84}}                        & \textbf{0.058}     & \multicolumn{1}{c|}{\textbf{7.5}}                         & \textbf{0.38}      \\ \hline
\end{tabular}
}
\end{table} 

% \usepackage{multirow}
\begin{table*}[t]
\centering
\caption{Speaker recognition evaluations on different test protocols for baseline systems and proposed wav2vec 2.0 based systems in terms of EER{[}\%{]} / minDCF(0.05)}

\label{tab:main_table}
\scalebox{0.84}{
\centerline{
\begin{tabular}{ccccccccc}
\hline
\rowcolor[HTML]{FFFFC7} 
\multicolumn{1}{|c|}{\cellcolor[HTML]{FFFFC7}{\color[HTML]{172B4D} }} &
  \multicolumn{1}{c|}{\cellcolor[HTML]{FFFFC7}} &
  \multicolumn{7}{c|}{\cellcolor[HTML]{FFFFC7}\textbf{Test datasets}} \\ \cline{3-9} 
\rowcolor[HTML]{FFFFC7} 
\multicolumn{1}{|c|}{\multirow{-2}{*}{\cellcolor[HTML]{FFFFC7}{\color[HTML]{172B4D} \textbf{System}}}} &
  \multicolumn{1}{c|}{\multirow{-2}{*}{\cellcolor[HTML]{FFFFC7}\textbf{\#Params, M}}} &
  \multicolumn{1}{c|}{\cellcolor[HTML]{FFFFC7}SRE'18 dev} &
  \multicolumn{1}{c|}{\cellcolor[HTML]{FFFFC7}SRE'16 eval} &
  \multicolumn{1}{c|}{\cellcolor[HTML]{FFFFC7}SRE'19 eval} &
  \multicolumn{1}{c|}{\cellcolor[HTML]{FFFFC7}VC1-O (cleaned)} &
  \multicolumn{1}{c|}{\cellcolor[HTML]{FFFFC7}VOiCES dev} &
  \multicolumn{1}{c|}{\cellcolor[HTML]{FFFFC7}CTS'20 progress $^1$} &
  \multicolumn{1}{c|}{\cellcolor[HTML]{FFFFC7}SRE'21 eval $^2$} \\ \hline
\multicolumn{9}{c}{\textit{\begin{tabular}[c]{@{}c@{}}Baseline encoders\end{tabular}}} \\ \hline \hline
\multicolumn{1}{|c|}{ResNet101} &
  \multicolumn{1}{c|}{27.5} &
  \multicolumn{1}{c|}{3.28/0.118} &
  \multicolumn{1}{c|}{5.01/0.237} &
  \multicolumn{1}{c|}{2.39/0.134} &
  \multicolumn{1}{c|}{1.78/0.105} &
  \multicolumn{1}{c|}{1.81/0.110} &
  \multicolumn{1}{c|}{2.75/0.097} &
  \multicolumn{1}{c|}{5.41/0.344} \\ \hline
\multicolumn{1}{|c|}{ECAPA-TDNN} &
  \multicolumn{1}{c|}{29} &
  \multicolumn{1}{c|}{4.14/0.152} &
  \multicolumn{1}{c|}{8.59/0.337} &
  \multicolumn{1}{c|}{2.97/0.165} &
  \multicolumn{1}{c|}{1.87/0.123} &
  \multicolumn{1}{c|}{2.02/0.123} &
  \multicolumn{1}{c|}{2.91/0.109} &
  \multicolumn{1}{c|}{6.26/0.398} \\ \hline
\multicolumn{1}{|c|}{\begin{tabular}[c]{@{}c@{}}ResNet101 \\ + ECAPA TDNN\end{tabular}} &
  \multicolumn{1}{c|}{56.5} &
  \multicolumn{1}{c|}{3.17/0.114} &
  \multicolumn{1}{c|}{4.87/0.221} &
  \multicolumn{1}{c|}{2.12/0/122} &
  \multicolumn{1}{c|}{1.35/0.086} &
  \multicolumn{1}{c|}{1.31/0.081} &
  \multicolumn{1}{c|}{2.71/0.085} &
  \multicolumn{1}{c|}{4.74/0.299} \\ \hline
\multicolumn{9}{c}{\textit{\begin{tabular}[c]{@{}c@{}}New encoders\end{tabular}}} \\ \hline \hline
\multicolumn{1}{|c|}{\begin{tabular}[c]{@{}c@{}}wav2vec-TDNN \\ (XLSR\_53)\end{tabular}} &
  \multicolumn{1}{c|}{98} &
  \multicolumn{1}{c|}{3.07/0.137} &
  \multicolumn{1}{c|}{4.18/0.206} &
  \multicolumn{1}{c|}{2.34/0.142} &
  \multicolumn{1}{c|}{0.82/0.052} &
  \multicolumn{1}{c|}{0.99/0.06} &
  \multicolumn{1}{c|}{\textbf{2.25/0.080}} &
  \multicolumn{1}{c|}{4.43/0.283} \\ \hline
\multicolumn{1}{|c|}{\begin{tabular}[c]{@{}c@{}}wav2vec-TDNN \\ (XLS-R\_1B)\end{tabular}} &
  \multicolumn{1}{c|}{265} &
  \multicolumn{1}{c|}{\textbf{2.94/0.083}} &
  \multicolumn{1}{c|}{\textbf{3.13/0.161}} &
  \multicolumn{1}{c|}{\textbf{1.71/0.097}} &
  \multicolumn{1}{c|}{\textbf{0.69/0.040}} &
  \multicolumn{1}{c|}{1.02/0.057} &
  \multicolumn{1}{c|}{3.61/0.080} &
  \multicolumn{1}{c|}{\textbf{3.59/0.281}} \\ \hline
\end{tabular}
}
}
\end{table*}

\subsection{S-norm settings}
\label{ssec:sn_cohort}
The s-norm cohort for score normalization consisted of the following sets: NIST SRE'18 unlabeled \cite{sre18}, NIST SRE'16 development and unlabeled sets \cite{sadjadi20172016},  IARPA Babel datasets \cite{babel}. 
We used top 200 scores to compute s-norm statistics. 

\subsection{Evaluation data and metrics}
\label{ssec:test_set}
The following sets were used for the evaluation:
\begin{itemize}
\setlength\itemsep{-0.3em}
    \item \textbf{Microphone sets}: VoxCeleb1-O (VC1-O) cleaned test set \cite{nagrani2017voxceleb}, VOiCES development set \cite{nandwana2019voices};
    \item \textbf{Telephone sets}: NIST SRE 2018 development set \cite{sre18}, NIST SRE 2016 evaluation set \cite{sadjadi20172016}, NIST SRE 2019 evaluation set \cite{sadjadi20202019},   NIST 2020 CTS progress \cite{sadjadi2020nist};
    \item \textbf{Cross channel set}: SRE 2021 challenges sets \cite{evalplan2021nist} .
\end{itemize}
We evaluate SR systems performance in terms of Equal Error Rates (EER) and minimum detection cost functions (minDCF) with $P_{tar}=0.01$ and $P_{tar}=0.05$ \cite{sadjadi20202019}.

\section{Results and discussion}
Tables \ref{tab:w2v-tdnnclean} and \ref{tab:w2v-tdnnaug} demonstrate the results of our preliminary experiments of wav2vec-TDNN(\textit{XLSR\_53})-based systems using clean and augmented versions of VC1 train set. The results were obtained for microphone VC1-O (cleaned) and telephone SRE'18 dev evaluation protocols. The performance of the system depending on wav2vec 2.0 transformer layer selection is shown on the Tables. It can be seen from the results of Table \ref{tab:w2v-tdnnaug} that there is no need to use the entire deep wav2vec 2.0 encoder architecture with 24 layers for such SR task. The 6\textsuperscript{th} encoder layer provides speaker recognition results comparable to other deeper transformer encoder layers. Thus in our further experiments with wav2vec-TDNN(\textit{XLSR\_53}) based systems we used 6\textsuperscript{th} layer transformer encoder network before TDNN block. For the  wav2vec-TDNN based on \textit{XLS-R\_1B} architecture, 12\textsuperscript{th} layer was chosen as optimal.
The online augmentations did not help to improve SR systems performance in the case of a small VC1 tuning set and 20 epochs tuning procedure.  We guess the reason for that is a lack of clean in-domain data which the network "has seen" during the training in this scenario.  

Table \ref{tab:w2v-aug} reveals the positive effect of using augmentation procedure during wav2vec-TDNN(\textit{XLSR\_53}) extractor fine-tuning for SR. One can see that data augmentation improves system robustness in telephone domain when only microphone data (VC1 and VC2) is used for training. 
Moreover, Tables \ref{tab:w2v-tdnnclean}, \ref{tab:w2v-tdnnaug} and \ref{tab:w2v-aug} show impressive and state-of-the-art results of the systems fine-tuned on relatively small sets VC1 (1211 speakers) and VC1+VC2 (7205 speakers).

For comparison of baseline and new wav2vec 2.0 based deep speaker embedding extractors performance on different evaluation protocols see Table  \ref{tab:main_table}. These results show the robustness of new encoders to different acoustic conditions in comparison to considered baseline systems. Our best and largest wav2vec-TDNN(\textit{XLS-R\_1B}) model demonstrates strong stability across microphone and telephone evaluation data achieving $EER = 0.69\%$ on VC1-O(cleaned) and $EER=1.71\%$ on SRE'19 eval protocols. One should note that there is a difference in baseline and wav2vec 2.0 models complexity in terms of the number of trainable parameters. According to our results increasing the model complexity for ResNet or ECAPA-TDNN did not lead to better robustness to different domains. We also tried to add SpecAugment for baseline systems training but did not observe any performance improvements.

Another thing we should note is that our attempts to train wav2vec-TDNN SR systems from scratch were unsuccessful. Thus we conclude that an unsupervised autoregressive pretraining scheme (for example with Contrastive Predictive Coding loss) efficiently utilizes the power of unlabeled data and opens the door to powerful transformer-based speaker embedding extractors. 

%7205 speakers:\\

\section{Conclusions}
Large transformer-based speaker embedding extractors can be developed with the help of unsupervised speech representation learning schemes. Our experiments for wav2vec 2.0 on a wide range of verification protocols reveal that such models are powerful.
Presented wav2vec-TDNN models fine-tuned on diverse training sets with augmentations demonstrate good robustness and generalization across different acoustic domains.

It was shown that fine-tuning of wav2vec-TDNN architectures for specific domains can be done on relatively small sets of data. Using data augmentation during fine-tuning provides additional performance gains in speaker verification.

\section{Acknowledgements}
We express our gratitude and deep appreciation to our colleagues from Speaker Recognition Team: T. Kotov, A. Kozlov, I. Korsunov, A. Vinogradova, A. Shulipa, T. Pekhovsky  and from Automatic Speech Recognition Team: I. Medennikov, M. Korenevsky, Y. Khokhlov, M. Korenevskaya, T. Prisyach and T. Timofeeva for the valuable advices and interesting discussions.
\footnotetext[1]{evaluated using NIST SRE platform }
\footnotetext[2]{evaluated using NIST SRE 2021 scoring tool}

\vfill\pagebreak
\bibliographystyle{IEEEtran}

% \bibliography{refs.bib}

\end{document}